\documentclass[a4paper,11pt]{article}
\usepackage{pos}

\title{Jet measurements at CMS}
\manuallySeparateAuthors
\author[a]{Cristian Baldenegro}
\author{ for the CMS Collaboration}

\affiliation[a]{The University of Kansas,\\
  Lawrence, Kansas, U.S.}

\emailAdd{c.baldenegro@cern.ch}

\abstract{Recent measurements based on jet production in high energy pp collisions at the CERN LHC with the CMS detector are reported. Specifically, the measurement of the inclusive jet production cross section as a function of the anti-$k_t$ distance parameter $R$ divided by the jet cross section at $R = 0.4$, is discussed. The cross section ratio is sensitive to various perturbative and non-perturbative treatments of the jet formation process in quantum chromodynamics (QCD). In the second study, the regimes of validity of the parton shower and matrix element approaches are tested on multijet events in pp collisions at $\sqrt{s} = 8$ and $13$ TeV. Special attention is given to the second and third leading $p_\text{T}$ jets and their momenta and rapidity--azimuth correlations. Finally, studies of dijet production where the two leading $p_T$ jets are separated by a large pseudorapidity interval void of charged particles is presented. This signature is expected from hard color singlet exchange (two-gluon exchange in perturbative QCD). The latter can be treated with perturbative QCD techniques based on the Balitsky--Fadin--Kuraev--Lipatov (BFKL) evolution equations.}

\FullConference{%
  40th International Conference on High Energy physics - ICHEP2020\\
  July 28 - August 6, 2020\\
  Prague, Czech Republic (virtual meeting)
}

\begin{document}
\maketitle

Quantum chromodynamics (QCD) is the gauge theory describing the strong interaction between quarks and gluons. Jets, the collimated sprays of hadron particles, approximate the properties of the original partons created in short-distance scatterings. The production cross sections for high transverse momentum ($p_\text{T}$) partons can be calculated using perturbative QCD (pQCD) techniques. Specifically, predictions for hadron production in proton-proton (pp) collisions require models for parton showering, and non-perturbative (NP) corrections, such as hadronization and underlying event (UE) activity. For certain observables and kinematic configurations, the fixed-order predictions from pQCD are not adequate, and thus higher-order terms must be accounted for using resummation methods. Studies of jet production can be used to further understand our modeling of strong interactions at short and long distances with great precision. In this report, we present a selected number of results by the CMS experiment~\cite{CMS} related to recent jet measurements.

The first study is related to inclusive jet cross section as a function of $R$ by CMS~\cite{Sirunyan:2020uoj}. Quarks and gluons radiate secondary gluons that can be emitted outside of the catchment area of the jet definition, which is the region in rapidity--azimuth plane contributing to the jet. This lost $p_\text{T}$ is calculated using a QCD splitting function embedded in parton shower algorithms, which are mostly important at small $R$. Properties of jets are also modified by hadronization, an NP process describing the transition of partons into hadrons, whose effects become more evident at larger $R$. In order to understand the underlying mechanisms at play contributing to the jet formation process, it is instructive to analyze the production of jets at various sizes of the anti-$k_t$~\cite{Cacciari_2008} parameter $R$ in a systematic way.

A measurement of the ratio of cross sections of inclusive anti-$k_t$ jets of multiple sizes with respect to jets with the distance parameter $R=0.4$ has been recently presented by CMS. Because of cancellation of many experimental and theoretical uncertainties in the ratio, one can achieve better sensitivity to perturbative and NP effects than that obtained with absolute cross section measurements. The analysis is based on data recorded in pp collisions in 2016, corresponding to an integrated luminosity of 35.9 fb$^{-1}$. The jets have $84 < p_\text{T} < 1588$ GeV and $|y|<2.5$ and are all clustered with the anti-$k_t$ algorithm. The main observable in this measurement is the double-differential inclusive jet cross section ratio, $ (d^2\sigma(R)/dp_T dy )\,/\, (d^2\sigma(R=0.4)/dp_Tdy) $.

\begin{figure}[]
\centering
\includegraphics[width=0.5\textwidth]{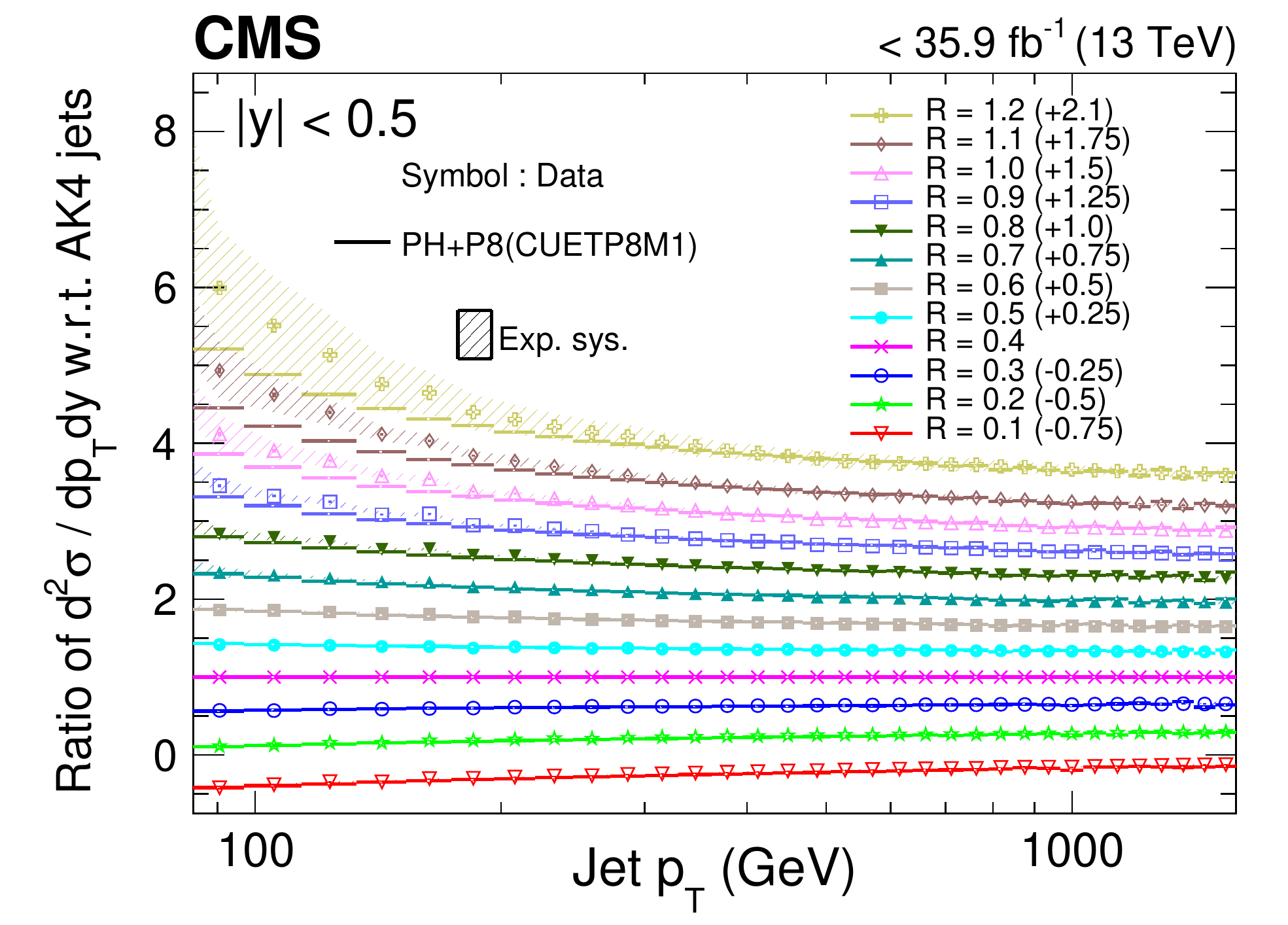}
\caption{\label{fig:ratio_xsec}Comparison of the ratio of the differential cross sections of jets of different sizes with respect to that of AK4 jets from data and from NLO predictions using \textsc{Powheg+Pythia8} (CUETP8M1 tune) in the region $|y|<0.5$. The distributions are artificially displaced for better visibility. The figure is extracted from Ref.~\cite{Sirunyan:2020uoj}.}
\end{figure}

The ratio is done in reference to $R = 0.4$ jets, since this is the standard $R$ value used for Run-II measurements by CMS. The cross section ratios are shown in Fig.~\ref{fig:ratio_xsec}. The results are compared to those generated by the NLO \textsc{Powheg} Monte Carlo (MC) generator~\cite{powheg} supplemented with \textsc{Pythia8} parton shower~\cite{pythia8}. A slope in the cross section ratios is observed at low $p_\text{T}$. The slope becomes more pronounced at larger $R$ values. This is consistent with the expected contributions of the underlying event activity. The cross section ratios are observed to flatten at higher $p_T \sim 200$--$300$ GeV for the probed $R$ values. The measured cross section ratios are also compared with predictions leading order (LO) and next-to-LO (NLO) fixed-order pQCD predictions, which are described in Ref.~\cite{Sirunyan:2020uoj}. A good agreement is achieved by NLO predictions over the $R$ parameter range probed in the measurement. The NLO predictions slightly overshoot the data at larger $R$ values for forward jets $1.5<|y|<2.0$, as shown in Ref.~\cite{Sirunyan:2020uoj}.

We now turn to the second study by CMS~\cite{CMS-PAS-SMP-17-008}, where the collinear and wide angle radiation regimes in different regions of the jet $p_\text{T}$ are studied. The understanding of the mechanisms responsible for multijet final states is crucial for our modeling of strong interactions. Two different scenarios are presented, one with three-jets and another one with Z+2 jet final states (not discussed in this report, but can be found in Ref.~\cite{CMS-PAS-SMP-17-008}). The first, second, and third leading $p_T$ jets are labeled as $j1$, $j2$, and $j3$, respectively. For one particular configurations, one could test the regimes of validity of matrix element approach (harder parton splittings, wide angle emission) and parton shower approaches (soft collinear emissions). One of the main variables is the angular correlation between the second and third jet, $\Delta R_{23} \equiv \sqrt{\Delta y_{23}^2+\Delta \phi_{23}^2}$, and the $p_{T3}/p_{T2}$ ratio distributions. The study is based on 8 TeV (20 fb$^{-1}$) and 13 TeV (2.3 fb$^{-1}$) data. In these events, the leading jet has $p_\text{T1}>510$ GeV. The rest of the jets have $p_T > 30$ GeV, and the first and second jets have $|y_{1,2}|<2.5$. The leading two jets have $2.14 < \Delta\phi_{12} < \pi$, to select for the trijet topology of interest. The second and third jets satisfy the ratio $0.1 < p_\text{T3}/p_\text{T2} < 0.9$ and angular separation $R_\text{jet}+0.1 < \Delta R_{23} < 1.5$.
 
\begin{figure}[]
\centering
\includegraphics[width=0.35\textwidth]{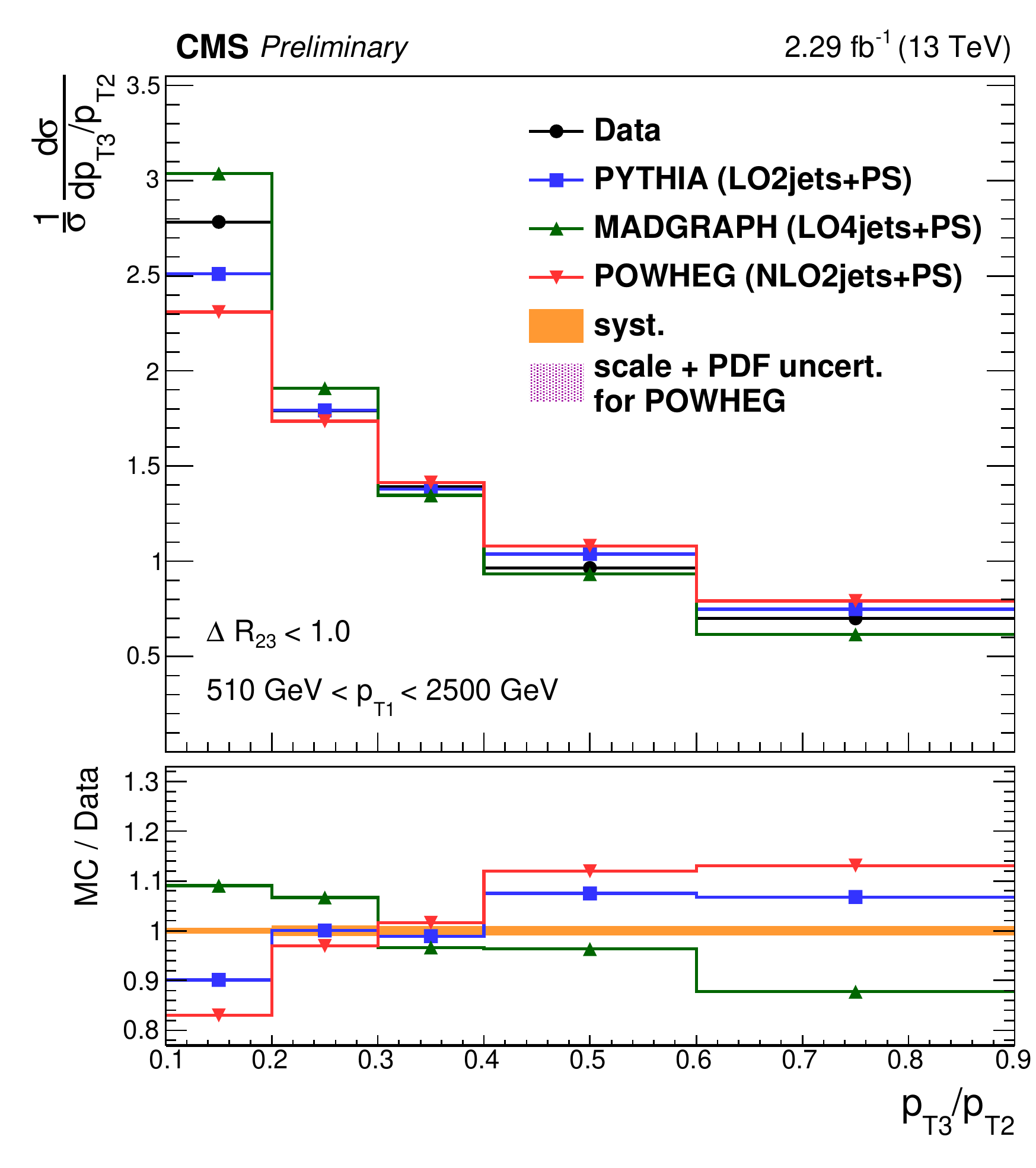}
\includegraphics[width=0.35\textwidth]{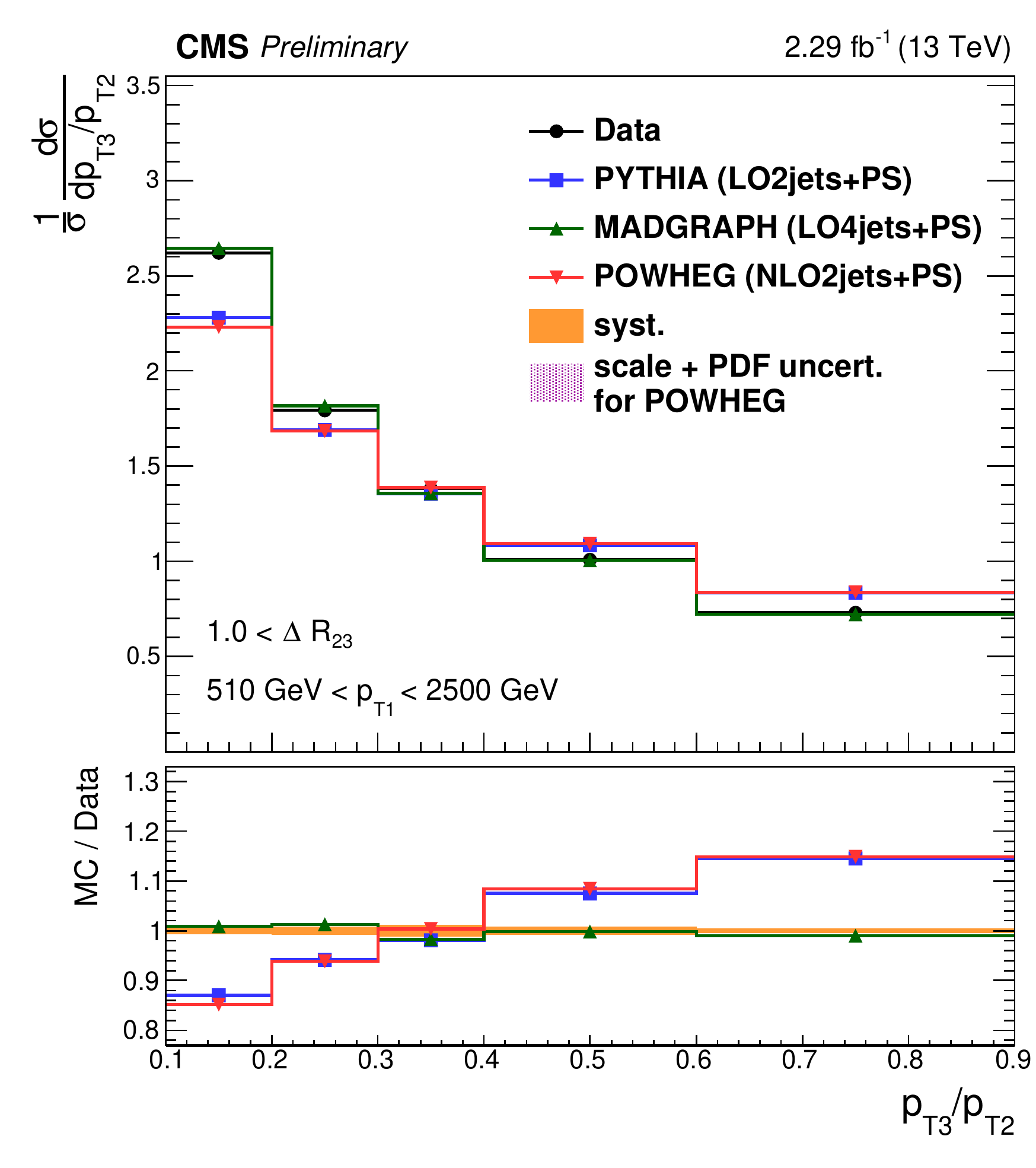}
\includegraphics[width=0.35\textwidth]{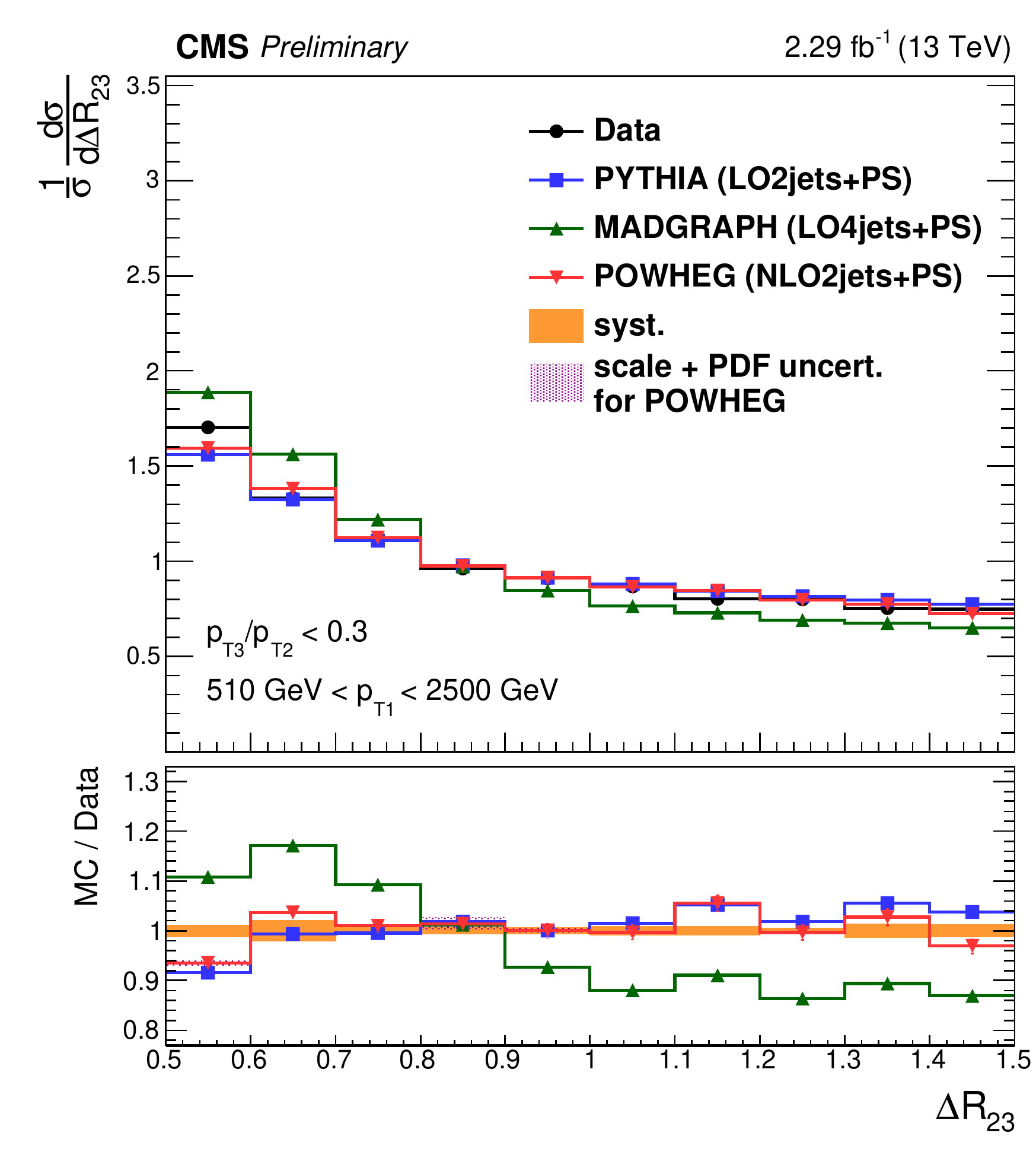}
\includegraphics[width=0.35\textwidth]{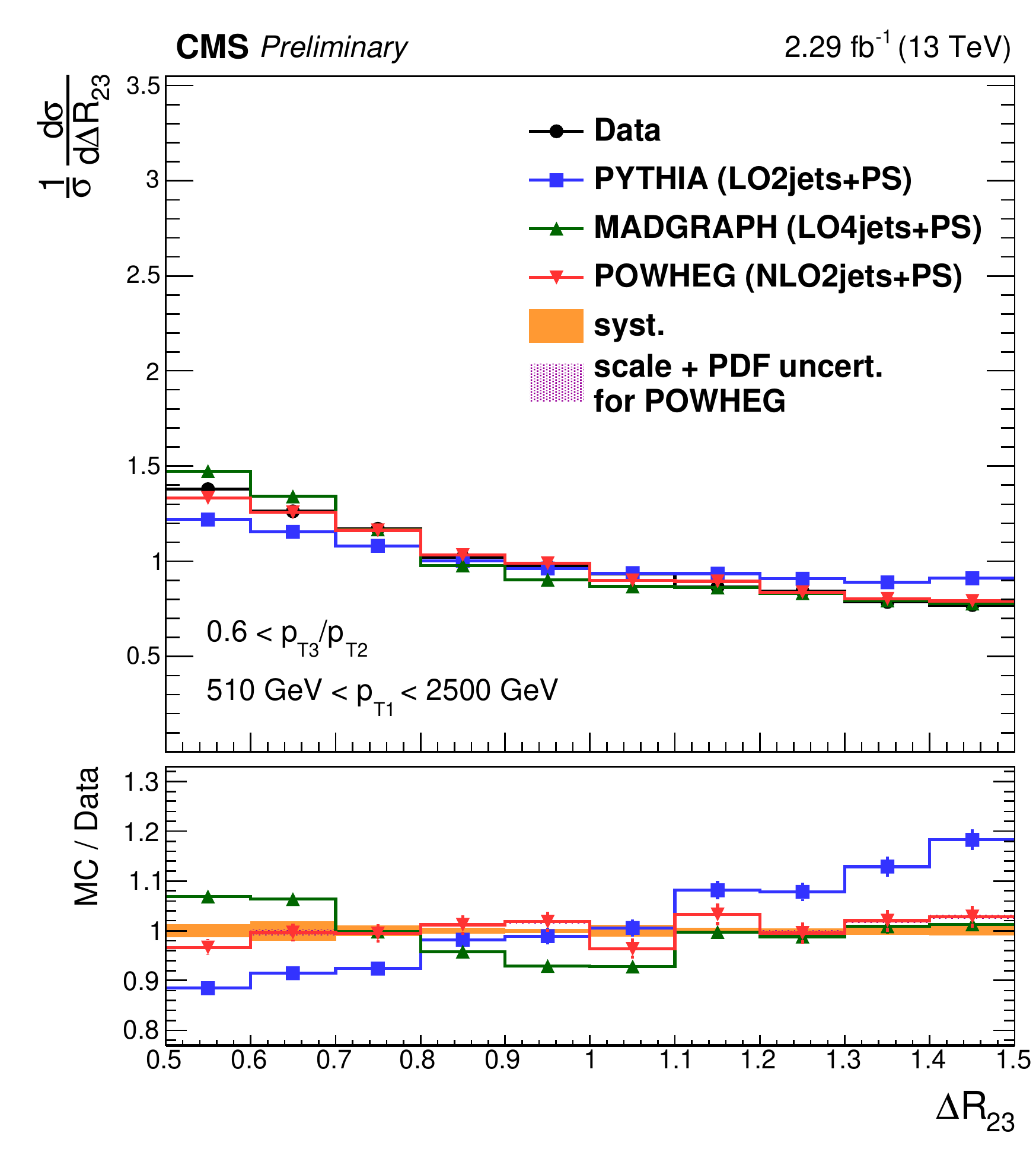}
\caption{\label{fig:dR_pTratio} (Upper left) $p_\text{T3}/p_\text{T2}$ for small angle radiation ( $\Delta R_{23} < 1.0$), (upper right) $p_\text{T3}/p_\text{T2}$ for large angle radiation ($\Delta R_{23} > 1.0$). (Lower left) $\Delta R_{23}$ for soft $p_\text{T}$ radiation ($p_\text{T3}/p_\text{T2}<0.3$), (lower right) $\Delta R_{23}$ for hard $p_\text{T}$ radiation ($p_\text{T3}/p_\text{T2}>0.6$). The data are compared to various theoretical predictions. The plots are extracted from Ref.~\cite{CMS-PAS-SMP-17-008}.}
\end{figure}

The experimental results for $\sqrt{s} = 13$ TeV are shown in Fig.~\ref{fig:dR_pTratio}. The data are compared with predictions based on \textsc{Powheg+Pythia8} (NLO+PS) and \textsc{Pythia8} (LO+PS), which are not able to describe data for small- and wide-angle emisisons in the soft- and hard-regimes in $p_{T3}/p_{T2}$. \textsc{MadGraph} (LO4jets + PS)~\cite{Alwall:2014hca} is able to describe $p_{T3}/p_{T2}$ spectra for wide-angle emissions $\Delta R_{23} > 1.0$. However, \textsc{MadGraph} does not describe well data for small-angle emissions $\Delta R_{23} <$ 1.0. Here, \textsc{Powheg} (NLO2jets+PS) gives the best agreement in $\Delta R_{23}$ for soft splittings ($p_\text{T3}/p_\text{T2} < 0.3$) and harder splittings ($p_\text{T3}/p_\text{T2} > 0.6$) regimes. Similar conclusions are drawn for the 8 TeV results (not shown in this report, but can be found in Ref.~\cite{jetgapjet_13TeV}).

The third and final study is related to hard color singlet exchange in dijet events studied by the CMS Collaboration~\cite{jetgapjet_13TeV}. In collisions with $t$-channel color singlet exchange between partons, the net color-flow is neutralized. In pQCD, this can be achieved with two-gluon $t$-channel exchange, where the second gluon screens the color charge of the first one. The experimental signature of this is a rapidity interval void of particle production between jets (rapidity gap). In the high-energy limit of QCD, color-singlet exchange corresponds to perturbative pomeron exchange. Thus, the process can be used to probe Balitsky--Fadin--Kuraev--Lipatov (BFKL) evolution, which resums diagrams that contribute to the amplitude with terms proportional to $\alpha_s^n \log^n(\hat{s}/|\hat{t}|) \sim \mathcal{O}(1)$ to all orders in $\alpha_s$, where $\hat{s}$ and $\hat{t}$ are the partonic center-of-mass energy and four-momentum transfer, respectively. The Dokshitzer-Gribov-Lipatov-Altarelli-Parisi (DGLAP) dynamics are strongly suppressed in jet-gap-jet events (Sudakov form factor for gap).

The analysis is based on low pileup 2015 data $\sqrt{s}= 13$ TeV~\cite{jetgapjet_13TeV}. For this investigation, anti-$k_t$ jets with $R = 0.4$ are used. The leading two jets should have $p_\text{T} > 40$ GeV and $1.4 < |\eta_\text{jet}|<4.7$ with $\eta_\text{jet-1} \times \eta_\text{jet-2} < 0$. The aforementioned $\eta$ selection requirements favor $t$-channel color singlet exchange to take place, and allows for a better separation of color exchange background events from the color singlet exchange signal events. In this study, the central pseudorapidity gap corresponds to absence of charged particle tracks between jets with $p_T^\text{ch}>200$ MeV and $|\eta|<1$. 

Standard dijet events, produced mostly by color exchange, dominate at high $N_\text{tracks}$ multiplicities. The high multiplicity region can be used to estimate the fluctuations at lower $N_\text{tracks}$ , where jets produced by color singlet exchange are expected to be present. Two data-based approaches are used to estimate the contribution of color exchange dijet events at low multiplicities. The first one consists of forming a second set of multiplicity distributions obtained from a sample of events where the two leading jets are on the same hemisphere of the detector ($\eta_\text{jet-1}\times \eta_\text{jet-2} > 0$, and the second consists on the use of a fit based on the negative binomial distribution (NBD) function, which are described in detail in Ref.~\cite{jetgapjet_13TeV}.

\begin{figure}[]
\centering
\includegraphics[width=0.3\textwidth]{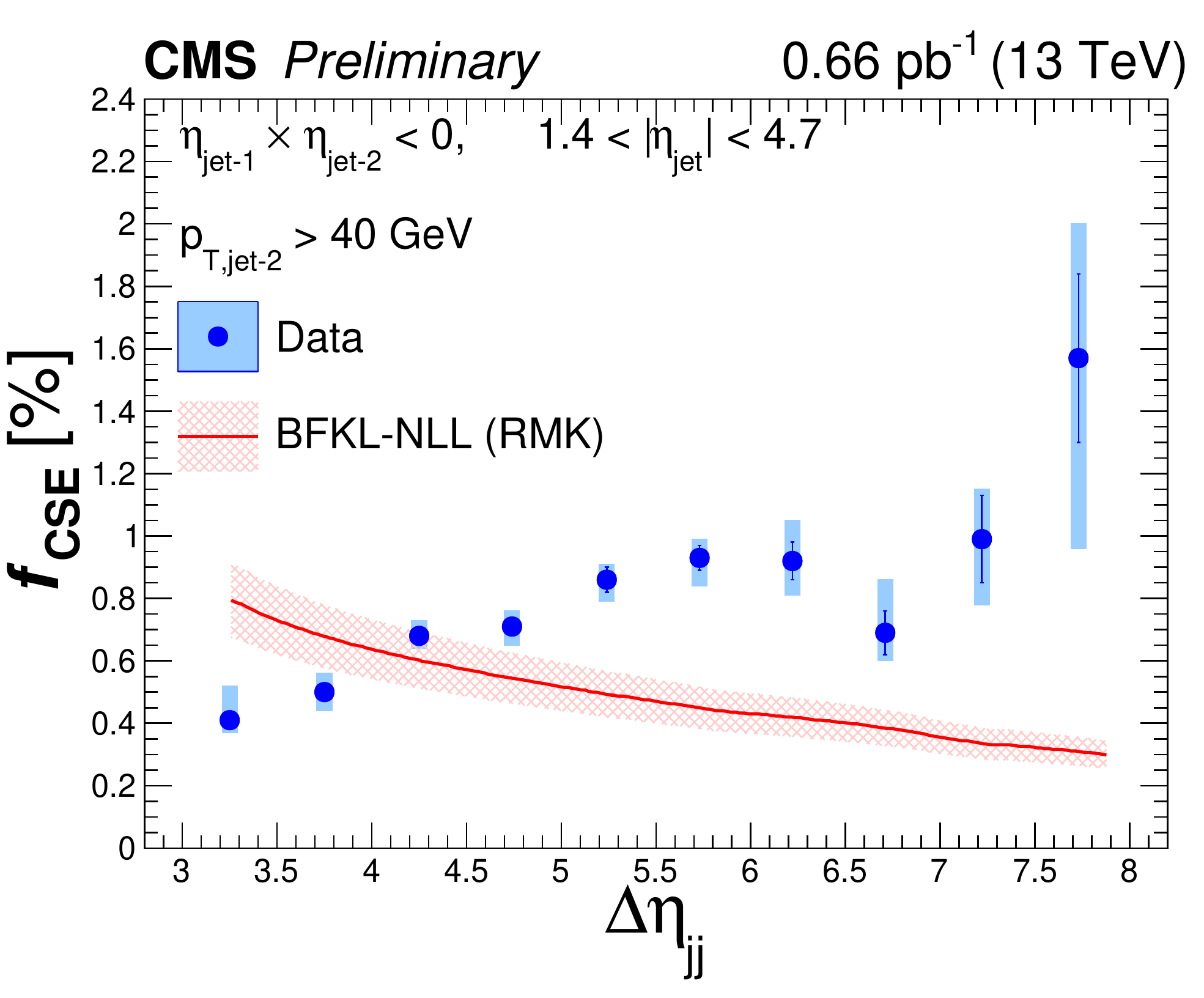}
\includegraphics[width=0.3\textwidth]{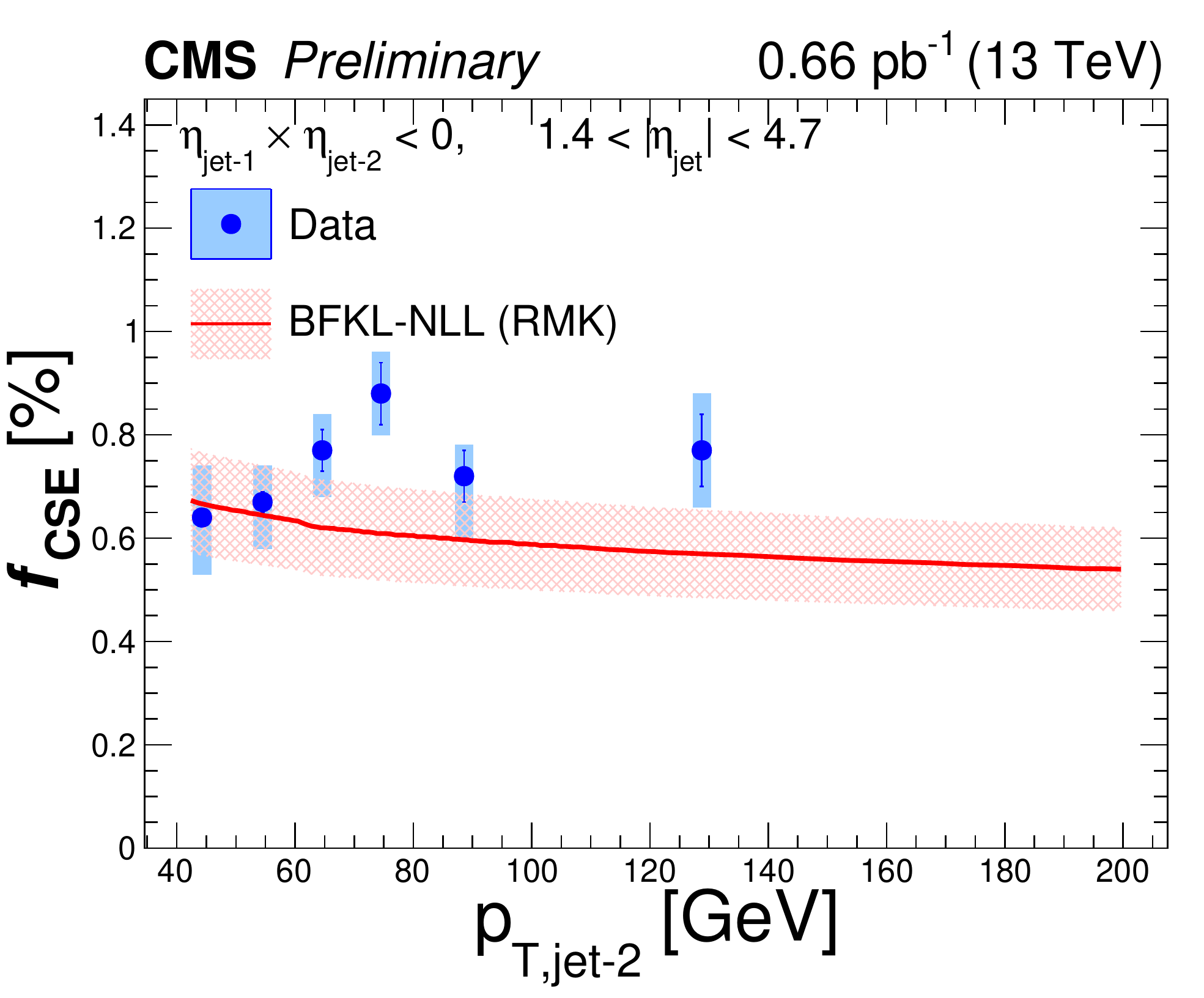}
\includegraphics[width=0.3\textwidth]{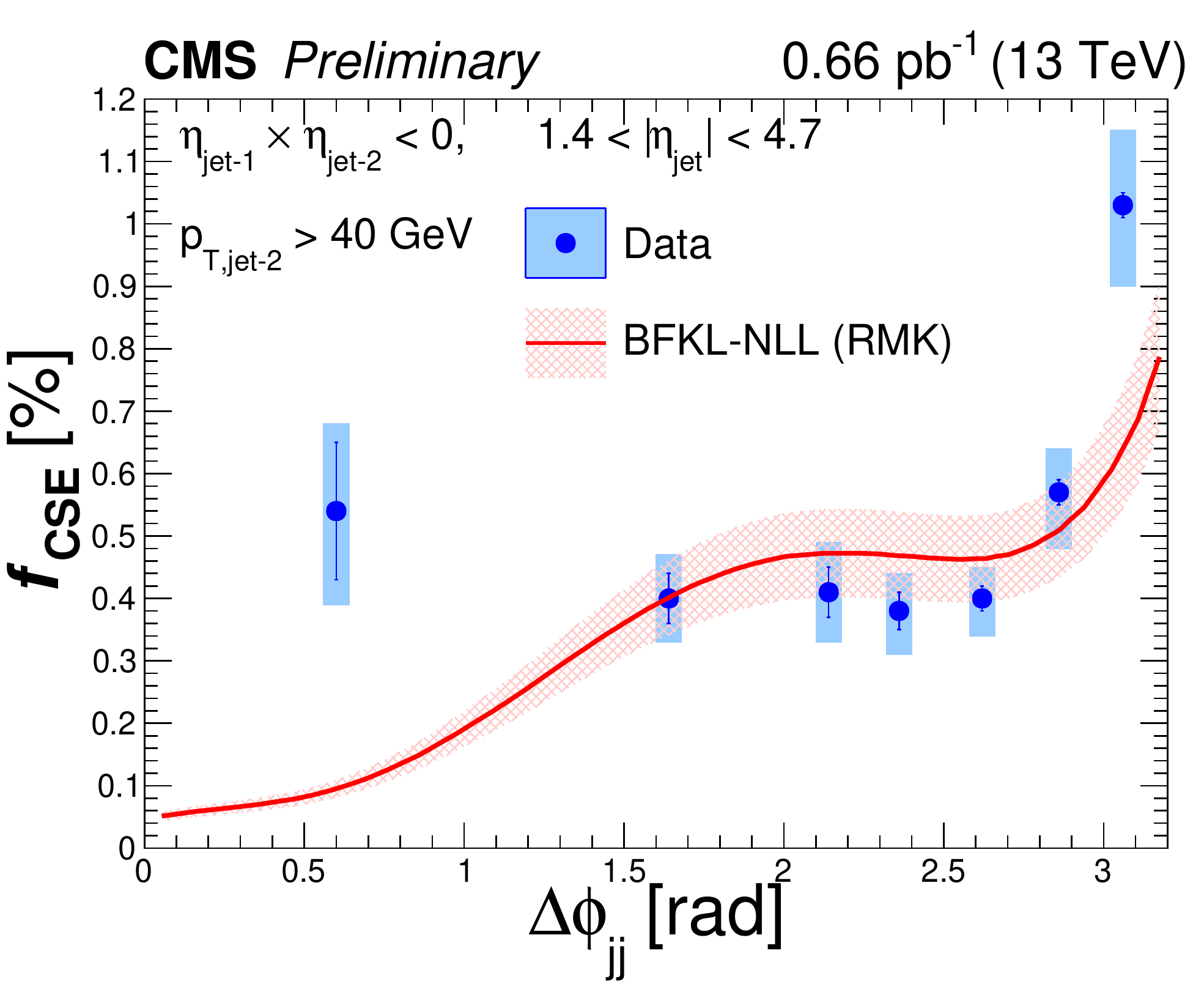}
\caption{\label{fig:fcse} Fraction of color singlet exchange dijet events, $f_\text{CSE}$, measured as a function of $\Delta\eta_\text{jj}$ , $p_\text{T,jet-2}$, and $\Delta\phi_\text{jj}$ in pp collisions at $\sqrt{s} = 13$ TeV. Vertical bars represent statistical uncertainties,
while boxes represent the combination of statistical and systematic uncertainties in quadrature. The figures are extracted from Ref.~\cite{jetgapjet_13TeV}.}
\end{figure}

The fraction of events produced by color singlet exchange, $f_\text{CSE}$, is extracted in this measurement. The $f_\text{CSE}$ generally increases with increasing $\Delta\eta_\text{jj} \equiv |\eta_\text{j1}-\eta_\text{j2}|$. A weak dependence of $f_\text{CSE}$ on $p_\text{T,j2}$ is observed. The $f_\text{CSE}$ increases at $\Delta\phi_\text{jj} \equiv |\phi_\text{j1}-\phi_\text{j2}| \approx \pi $, and is uniform at $\Delta\phi_\text{jj} < 2.7$. Typical values of $f_\text{CSE} = 0.6$--$1.0$\%. The data are compared to predictions by Royon, Marquet, Kepka (RMK) model based on BFKL calculations at next-to-leading logarithmic accuracy supplemented with LO impact factors, and a survival probability of $|S|^2 = 0.1$ used to match the data. As shown in Fig.~\ref{fig:fcse}, it is very challenging to describe all features of the measurement simultaneously. The measurement, and current disagreement of data with theory, provides further guidance for subsequent theory development.

To summarize, a number of recent jet-related measurements for studies of strong interactions have been presented. The respective measurements are sensitive to different theory approaches based on perturbative and non-perturbative methods in different kinematic regimes of (multi)jet production in proton-proton collisions at 13 TeV. The measurement of the inclusive jet cross section ratio relative to $R = 0.4$ is sensitive to various stages of the jet formation process. The ratio of cross sections with varying $R$ is well described by calculations that include corrections given by parton shower, but not by a leading-order (LO) quantum chromodynamics (QCD) calculation including non-perturbative effects. The next-to-LO (NLO) predictions, when supplemented with parton shower and non-perturbative corrections, give the best agreement with data for most of the parameter range. In the second investigation, where the focus is on multijet production, allows for the test of the regimes of validity of the matrix element and parton shower approaches. Special attention is given to the second and third leading $p_T$ jets angular and momentum correlations. The measurement indicates quantitatively where the methods of merging matrix element with parton shower calculations break down in particular regions of phase space. Finally, dijet events where the presence of charged particles with $p_T > 200$ MeV in $|\eta|<1$ between the jets is suppressed, as expected from hard color singlet exchange, are analyzed. Present predictions based on the Balitsky--Fadin--Kuraev--Lipatov framework (resummation of large logarithms of energy in perturbative QCD) are not able to describe all features of the data. The disagreement between the BFKL-based predictions and the data provide further guidance for model improvement.

\section*{Acknowledgments}

CB acknowledges the U.S. Department of Energy's grant DE-SC0019389 for the financial support provided in the investigation of jets separated by a pseudorapidity gap.

\end{document}